\documentclass{emulateapj}
\usepackage{apjfonts}
\submitted{}

\def\hi{\ifmmode {\rm H}\,{\sc i}~ \else H\,{\sc i}~\fi}

\def\chandra {{\it Chandra}}
\def\fuse {{\it FUSE}}
\def\cv {C\,\hbox{{\sc v}}}
\def\cvi {C\,\hbox{{\sc vi}}}
\def\nvi {N\,\hbox{{\sc vi}}}
\def\nvii {N\,\hbox{{\sc vii}}}
\def\neix {Ne\,\hbox{{\sc ix}}}
\def\ovi {O\,\hbox{{\sc vi}}}
\def\ovia {O\,\hbox{{\sc vi}}$_{\rm 1}$}
\def\ovib {O\,\hbox{{\sc vi}}$_{\rm 2}$}
\def\hvca {O\,\hbox{{\sc vi}}$_{\rm HVC1}$}
\def\hvcb {O\,\hbox{{\sc vi}}$_{\rm HVC2}$}
\def\ovii {O\,\hbox{{\sc vii}}}
\def\oviii {O\,\hbox{{\sc viii}}}

\def\novi {N_{\rm OVI}}
\def\novii {N_{\rm OVII}}
\def\noviii {N_{\rm OVIII}}
\def\pks {PKS~2155--304}
\def\kms    {\,km\,s$^{-1}$}

\slugcomment{Accepted to ApJ}

\shorttitle{Local Warm-Hot Gas Toward PKS 2155--304}
\shortauthors{Williams et al.}

\begin{document}


\title{{\it Chandra} and {\it Far Ultraviolet Spectroscopic Explorer} 
        Observations of $z\sim 0$ Warm--Hot Gas Toward \pks}

\author{Rik J. Williams\altaffilmark{1,2},
        Smita Mathur\altaffilmark{1},
	Fabrizio Nicastro\altaffilmark{3,4,5},
	Martin Elvis\altaffilmark{3}}
\altaffiltext{1}{Department of Astronomy, The Ohio State University, 
                 140 West 18th Avenue, Columbus OH 43210, USA}
\altaffiltext{2}{Current Affiliation: Leiden Observatory, Leiden University, 
                PO Box 9513, 2300 RA Leiden, The Netherlands}
\altaffiltext{3}{Harvard--Smithsonian Center for Astrophysics, 60 Garden Street,
		Cambridge, MA 01238, USA}
\altaffiltext{4}{Instituto de Astronom\'ia Universidad Aut\'onomica de
 M\'exico, Apartado Postal 70-264, Ciudad Universitaria, M\'exico, D.F.,
 CP 04510, M\'exico}
\altaffiltext{5}{Osservatorio Astronomico di Roma, Istituto Nazionale
di AstroFisica, Via di Frascati 33, I-00040 Monte Porzio Catone, Italy}
\email{williams@strw.leidenuniv.nl}

\begin{abstract}
The X-ray bright $z=0.116$ quasar \pks\ is frequently observed as
a \chandra\ calibration source, with a total of 483 ksec of Low Energy 
Transmission Grating (LETG) exposure time
accumulated through May 2006.  Highly--ionized metal 
absorption lines, including numerous lines at $z=0$ and a putative \oviii\
K$\alpha$ line at $z=0.055$, have been reported in past \chandra\ 
studies of this source.  Using all available \chandra\ LETG spectra and 
analysis techniques developed for such $z=0$ X-ray absorption along other 
sightlines, we revisit these previous detections.  We detect 4 absorption
lines at $>3\sigma$ significance (\ovii\ K$\alpha/\beta$, \oviii\ K$\alpha$,
and \neix\ K$\alpha$), with \ovii\ K$\alpha$ being a $7.3\sigma$ detection.
The $1\sigma$ ranges
of $z=0$ \ovii\ column density and Doppler parameter are consistent with those
derived for Mrk 421 and within $2\sigma$ of the Mrk 279 absorption.  
Temperatures and densities inferred from the relative \ovii\ and
other ionic column densities are found to be consistent with 
either the local warm--hot intergalactic medium or a Galactic corona.  Unlike 
the local 
X-ray absorbers seen in other sightlines, a link with
the low-- or high--velocity far--ultraviolet \ovi\ absorption lines cannot
be ruled out.  The $z=0.055$ \oviii\ absorption reported by Fang et al.~is
seen with $3.5\sigma$ confidence in the ACIS/LETG spectrum, but no other 
absorption lines are found at the same redshift.  

\end{abstract}


\section{Introduction}
At all known epochs, the baryonic mass density of the intergalactic medium
(IGM) is thought to outweigh the baryons found in denser collapsed objects-- 
stars, galaxies, and clusters.  Indeed, at high redshifts
($z\ga 2$) the ``forest'' of Ly$\alpha$ absorption lines seen in spectra of
distant quasars reveals a vast network of cool, photoionized hydrogen
that is consistent with the expected baryon density at those redshifts
\citep{weinberg97}.
More recently, however, the process of structure formation has
shock--heated this intergalactic gas to produce the warm--hot IGM 
\citep[WHIM;][]{cen99,dave01}.
At such high temperatures ($T\sim 10^5-10^7$\,K) and low densities
($10^{-6}-10^{-4}$\,cm$^{-3}$; $\delta\sim 5-500$), the combination of 
collisional-- and
photo--ionization renders most of the gas too highly ionized to be detected 
through its Ly$\alpha$ absorption \citep[though some broad Ly$\alpha$ systems
likely tracing the low-$z$ WHIM have been reported, 
e.g.][]{sembach04,richter04}.

As the strong Ly$\alpha$ transition is largely suppressed, most of the WHIM 
has proved extremely difficult to detect, resulting in a serious discrepancy
between the observed low-$z$ baryon census and predictions 
\citep[e.g., WMAP;][]{bennett03}.  However, at these temperatures and
densities heavier elements are \emph{not} fully ionized, leading to an
analogous ``forest'' of inner--shell X-ray metal absorption lines 
\citep{shapiro80,hellsten98,perna98}.  Even the most common heavy elements 
such as
oxygen and neon are $\sim 3-4$ orders of magnitude less abundant than hydrogen,
and only with the advent of \chandra's high--resolution X-ray spectroscopic
capabilities have such WHIM lines been directly observable toward bright,
low--redshift AGN
\citep[][Nicastro et al.~2007, in preparation]{fang02,nicastro05a,nicastro05b}.
Though the statistical errors are large with so few detections, the 
intervening X-ray absorption lines seen by \chandra\ are consistent
with the expected low-$z$ baryon density.

In addition to these intervening absorption lines, similar metal lines
at velocities consistent with zero are seen in several high--quality 
\chandra\ spectra \citep{nicastro02,nicastro03,williams05,williams06a}.
Since very little information about the physical distribution of this
absorbing medium is available, it is unknown whether it originates in the
Galaxy (e.g., as part of a hot, low--density corona) or farther away
in the local WHIM filament or local intra--group medium.  While there is
some evidence for such a Galactic corona 
\citep[e.g.][]{sembach03,collins05,wang05},
detailed curve--of--growth and ionization balance analyses of the X-ray
absorption lines have found that they are also consistent with the 
temperatures and densities expected in the WHIM 
\citep{williams05, williams06a}.

The ubiquitous \ovi\ high--velocity clouds (HVCs) observed by \fuse\ 
\citep{wakker03} present a similar puzzle.
These clouds are typically seen as absorption lines with velocities 
inconsistent with Galactic rotation ($\left|v\right|\ga 100$\kms) and,
lacking distance information, it is unclear whether they are part of the
extended Galactic baryon distribution or local intergalactic medium.
Some are almost certainly associated with nearby \hi\ HVCs at similar 
velocities, such as Complex C and the Magellanic Stream, but others appear
completely isolated \citep{sembachetal03}.  These isolated \ovi\ HVCs
have an average velocity vector that is large in the local standard of
rest but minimized by transforming to the Local Group rest frame
\citep{nicastro03}, indicating that they may indeed be extragalactic.

The nature of the local X-ray absorption and \ovi\ HVCs (and the association
between the two) is thus an important consideration for the fields
of galaxy formation and cosmology.  In the Mrk 421 and Mrk 279 sightlines 
analyzed by \citet{williams05} and \citet{williams06a} respectively,
the $z=0$ X-ray absorption was found to be unassociated with any single
component of the \ovi\ absorption systems seen with \fuse\ along the
same sightlines.  However, these systems appeared quite different from each 
other; for example, the inferred velocity dispersion of the \ovii\ absorption
was relatively low toward Mrk 421 ($b\sim 40$\kms) and high toward
Mrk 279 ($b>77$\kms).  Mrk 279 also exhibited a far stronger \ovi\ 
HVC than Mrk 421, possibly related to the nearby presence of an \hi\ 
HVC (Complex C) at the same velocity.  With only two sightlines
analyzed to this level of detail thus far, comparable data in other
directions are crucial to assemble the overall picture of local
warm--hot absorption.

Here we present a detailed analysis of the \chandra\ Low--Energy
Transmission Grating observations of \pks, a bright BL Lac object at
$z=0.116$.  Subsets of these data have been previously presented by 
\citet{fang02}, who reported the detection of a $z=0.055$ \oviii\ 
absorption line from the WHIM, and \citet{nicastro02}, who focused
on the properties of the $z=0$ absorption.  Much more calibration
data have been made publicly available in the years since,
and when combined represent the second highest--quality
\chandra\ grating spectrum of an extragalactic source in terms of
counts per resolution element, and the last sightline currently
in the \chandra\ archive 
for which potentially extragalactic, $z=0$ X-ray absorption lines are
likely to be detected (excluding 3C 273, which lies in the direction of a 
supernova remnant).  Since the physical properties of the $z=0$ absorption
toward Mrk 421 and Mrk 279 differed substantially in some ways (velocity
dispersion and offset from the \ovi\ HVCs), a third sightline provides
valuable insight into the global properties of this absorption.

\begin{figure*}
\plotone{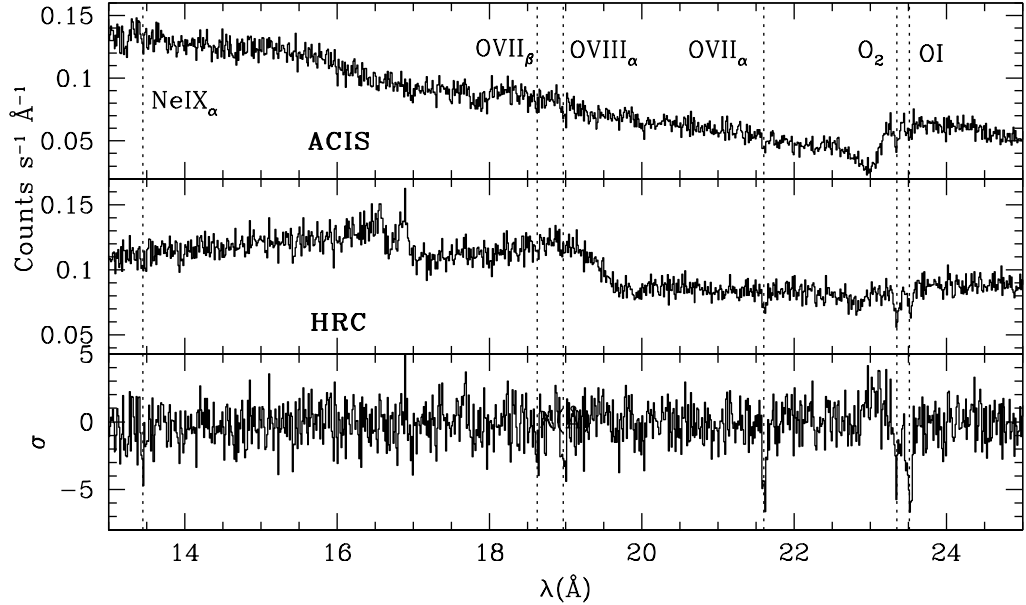}
\vspace{-6cm}
\caption{13--25\AA\ portion of the \chandra\ LETG grating spectra from 
ACIS-S (top panel)
and HRC-S (center).  The lower panel shows the sum of residuals from
both instruments after continuum fitting; detected $z=0$ absorption lines
are marked.  The neutral atomic and molecular oxygen lines near 23.5\AA\ are 
due primarily to the local interstellar medium and contaminants on the 
\chandra\ detectors.
\label{ch4_fig_bigacisspec}}
\end{figure*}

\section{Data Reduction and Measurements}
\subsection{\chandra}
\pks\ has been observed numerous times for calibration and science 
purposes with all possible combinations of the \chandra\ gratings and
detectors.  As the strongest absorption lines previously observed
have been lines from C, N, and O at $\lambda \ga 18$\,\AA, for the 
purposes of this study we only include data from the Low Energy
Transmission Grating (LETG) since it has the highest effective area
in this wavelength regime.  \chandra's two X-ray cameras, the High Resolution
Camera (HRC) and Advanced CCD Imaging Camera (ACIS), each include 
separate detector arrays for imaging (I) and grating spectroscopy (S).  
Although LETG observations taken with HRC--I and ACIS--I are available in 
the archive, their calibration is less certain and wavelength range more
restricted than those of the spectroscopic arrays, and so they are excluded
from this analysis.

The remaining datasets include 8 employing the HRC--S/LETG instruments and
24 with 
ACIS--S/LETG.  Of these latter observations, however, 15 have large pointing
offsets (typically $6^\prime-14^\prime$), presumably intended to characterize
the off--axis line spread function and effective area.  Since the spectral
resolution degrades significantly at these large offsets, only the nine
ACIS--S observations with $\left|\Delta\theta\right| \le 1\farcm 5$ are 
considered here.  The resulting 17 observations, listed in 
Table~\ref{ch4_tab_obs}, contain a total of
483\,ks of exposure time and $\sim 2100$ counts per 0.05\,\AA\ resolution
element (CPRE) at 21.5\,\AA\ (253.7\,ks 
and 880 CPRE in ACIS--S; 229.3\,ks and 1200 CPRE in HRC--S).  In theory
this should provide better than half the signal--to--noise ratio obtained in
the LETG spectrum of Mrk 421 taken during two
outburst phases \citep[6000 CPRE;][]{nicastro05a,williams05}.

All datasets were fully reprocessed using the Chandra Interactive Analysis of 
Observations (CIAO) software, version 3.3, with the corresponding
Calibration Database (CALDB) version 3.2.1\footnote{See 
\url{http://cxc.harvard.edu/ciao/} and \url{http://cxc.harvard.edu/caldb/}}.  
This CALDB version includes
models for the ACIS--S time--dependent quantum efficiency degradation as
well as preliminary corrections to nonlinearities in the HRC--S/LETG 
wavelength scale.  First--order spectra were then extracted, and response
matrices built, using the standard CIAO routines.  LETG spectral orders cannot
be separated with HRC--S due to this detector's intrinsic lack of energy
resolution, so the resulting spectrum is an overlapping superposition of 
all orders.  We thus built all HRC--S/LETG response matrices for orders
$-6$ to $+6$; past experience \citep[e.g.,][]{williams06a} has shown that this
is sufficient to accurately model higher--order contamination.

\begin{deluxetable}{lcccc}
\tablecolumns{5}
\tablecaption{\chandra\ observation log \label{ch4_tab_obs}}
\tablehead{
\colhead{Obs ID} &
\colhead{Date}  &
\colhead{$t_{\rm exp}$} &
\colhead{$f_\lambda$(21\AA)\tablenotemark{a}} &
\colhead{$W_i$\tablenotemark{b}}
\\
\colhead{} &
\colhead{} &
\colhead{(ks)} &
\colhead{(ks$^{-1}$\,cm$^{-2}$\,\AA$^{-1}$)} &
\colhead{}
}
\startdata
ACIS--S/LETG & & & & \\
1703 &2000 May 31 &25.2 &3.4 &0.121\\
2335 &2000 Dec 06 &29.1 &2.4 &0.090\\
3168 &2001 Nov 30 &28.8 &6.7 &0.223\\
3668 &2002 Jun 11 &13.5 &6.9 &0.120\\
3707 &2002 Nov 30 &26.9 &1.6 &0.054\\
4416 &2003 Dec 16 &46.5 &3.2 &0.174\\
6090 &2005 May 25 &27.5 &3.2 &0.106\\
6091 &2005 Sep 19 &29.2 &2.6 &0.087\\
6927 &2006 Apr 02 &27.0 &0.8 &0.025\\
\hline
HRC--S/LETG & & & & \\
331  &1999 Dec 25 &62.7 &9.7 &0.514\\
1013 &2001 Apr 06 &26.6 &2.9 &0.065\\
1704 &2000 May 31 &25.8 &3.7 &0.081\\
3166 &2001 Nov 30 &29.8 &8.4 &0.212\\
3709 &2002 Nov 30 &13.7 &2.3 &0.026\\
4406 &2002 Nov 30 &13.9 &2.4 &0.028\\
5172 &2004 Nov 22 &26.9 &1.8 &0.041\\
6923 &2006 May 01 &29.9 &1.3 &0.032\\
\enddata
\tablenotetext{a}{Background--subtracted photon flux at 21\AA; the HRC--S
values are apparent fluxes including all higher orders.  }
\tablenotetext{b}{Weight factors for coadding the response matrices, 
calculated as 
$W_i=f_{\lambda,i}(21\AA)t_{{\rm exp},i}/\Sigma_i(f_{\lambda,i}(21\AA)t_{{\rm exp},i})$.}
\end{deluxetable}

The individual spectra from each instrument were then coadded, both to
allow searches by eye for weak absorption lines and to make it
easier to assess the goodness of fits.  First, the 
positive and negative spectral orders from each observation (and their
corresponding response matrices) were coadded.  As \pks\ is highly variable 
in the X-ray band, we performed quick fits to determine the flux near 21\,\AA\
at the time of each observation.  The response matrices were then weighted
by a factor of $f_\lambda({\rm 21\AA})\times t_{\rm exp}$ and coadded; 
these fluxes and
weights are also listed in Table~\ref{ch4_tab_obs} and show a factor 12
variation at 21\AA\ from 1999 to 2006.  Note that, as a result of
the dithering strategy employed during observations, \chandra/LETG lacks the
narrow chip--gaps and other detector features seen in \emph{XMM--Newton}
grating spectra \citep{williams06b}; weighting the response matrices before
coaddition substantially reduces broad residuals across the LETG band, but 
is not essential for narrow absorption line measurements.

The resulting spectra were fit using the CIAO \emph{Sherpa} utility.
Simple powerlaw continua (with foreground Galactic absorption as a free
parameter) were fit independently for the ACIS and HRC spectra over the
10--47\,\AA\ wavelength range.  To improve the consistency of the fit near
elemental edges, the foreground absorber abundances of carbon, nitrogen, 
oxygen, and neon were allowed to vary.  The resulting best--fit powerlaw 
slopes are similar ($\Gamma=-0.63$ and $-0.45$ for ACIS and HRC respectively,
where $f_\lambda\sim \lambda^{-\Gamma}$), but the Galactic absorption
and abundances vary somewhat between the two instruments, perhaps due to
calibration uncertainties or a degeneracy between $\Gamma$ and 
$N_H$ over this restricted wavelength range.  The continuum fits for
ACIS and HRC are shown in Figure~\ref{ch4_fig_bigacisspec}.

\begin{figure}
\plotone{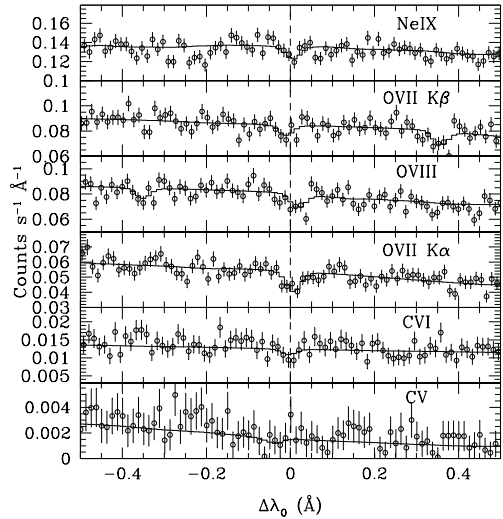}
\caption{ACIS-S/LETG data (points) and best--fit models (histogram) 
near each of the six $z=0$ X-ray absorption lines detected toward \pks.
\label{ch4_fig_acisspec}}
\end{figure}

\begin{figure}
\plotone{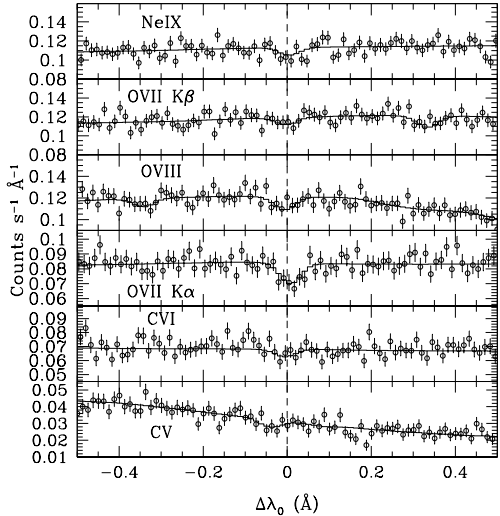}
\caption{Detected $z=0$ absorption lines in HRC--S; see 
Figure~\ref{ch4_fig_acisspec} for details.
\label{ch4_fig_hrcspec}}
\end{figure}

Several absorption lines, including the \ovii, \oviii, and \neix\ K$\alpha$ 
transitions at $z=0$, are immediately visible.
These lines were modeled with narrow (FWHM$<50$\,m\AA) Gaussian 
features added to the fitted continua.  Since there appear to be
lingering systematic uncertainties in the HRC--S/LETG wavelength scale
even with the new correction routines \citep{nicastro07},
the wavelengths and strengths of 
absorption features (as well as the continuum normalizations) were first 
allowed to vary independently for the
spectrum produced by each instrument.  Each line's equivalent width was
then determined using a joint fit to the ACIS and HRC spectra with the
requirement that the equivalent width match between the two instruments,
i.e.  the normalized Gaussian line amplitudes $A_{\rm HRC}$ and $A_{\rm ACIS}$ 
(where $A$ corresponds to the integral of the Gaussian, not the height) were 
fixed according to:
\begin{equation}
f_{\lambda,{\rm HRC}} A_{\rm HRC} = f_{\lambda,{\rm ACIS}} A_{\rm ACIS}
\end{equation}
Wavelength and equivalent width errors were determined for this joint
fit using the ``projection'' command in \emph{Sherpa}, allowing the HRC 
and ACIS continuum normalizations to vary.  These quantities for all
measured $z=0$ lines (as well as upper limits on \neix\ and \ovii\ K$\gamma$)
are reported in Table~\ref{ch4_tab_lines}, and Figures~\ref{ch4_fig_acisspec}
and \ref{ch4_fig_hrcspec} show the best--fit models for all detected lines
in ACIS and HRC respectively.  To check the consistency between the two
detectors, absorption lines were also fit independently in both spectra.
The resulting wavelengths and equivalent widths agree to within $1\sigma$,
with the exception of \neix\ which exhibited a $1.5\sigma$ (3.5\,m\AA) 
larger equivalent width in ACIS than in HRC. This could plausibly be due
to statistical fluctuations, however, so we conclude that the two detectors
produce consistent results.

\begin{deluxetable*}{lcccccccc}
\tablecolumns{9}
\tablecaption{Observed $z\sim 0$ absorption lines \label{ch4_tab_lines}}
\tablehead{
\colhead{ID} &
\colhead{$\lambda_{\rm rest}$\tablenotemark{a}} &
\colhead{$\lambda_{\rm obs}$\tablenotemark{b}} &
\colhead{$\Delta v_{\rm FWHM}$} &
\colhead{$v_{\rm obs}$} &
\colhead{$W_\lambda$\tablenotemark{c}} &
\colhead{$\log N_i$\tablenotemark{c,d}} &
\colhead{Significance} &
\colhead{Note}
\\
\colhead{} &
\colhead{(\AA)} &
\colhead{(\AA)} &
\colhead{(\kms)} &
\colhead{(\kms)} &
\colhead{(m\AA)} &
\colhead{} &
\colhead{$\sigma$} &
\colhead{}
}

\startdata
X-ray (\chandra): \\
\hline\hline
\ion{C}{5} K$\alpha$ &40.268 &$40.227^{+?}_{-.015}$ &\nodata &$-305^{+?}_{-112}$ &$11.4\pm 5.1$ & $15.22^{+0.26}_{-0.33}$ &2.2 &1\\
\ion{C}{6} K$\alpha$ &33.736 &$33.732^{+.011}_{-.007}$ &\nodata &$-36^{+98}_{-62}$ &$5.6\pm 2.5$ & $15.16^{+0.18}_{-0.27}$ &2.2 &\\
\ion{O}{7} K$\alpha$ &21.602 &$21.611^{+.002}_{-.008}$ &\nodata &$125^{+28}_{-111}$ &$11.6\pm 1.6$ & $16.09\pm 0.19$ &7.3 &\\
\ion{O}{7} K$\beta$  &18.629 &$18.618\pm .007$         &\nodata &$-177\pm 113$ &$4.2\pm 1.3$ & $16.09^{+0.17}_{-0.21}$ &3.2 &\\
\ion{O}{8} K$\alpha$ &18.969 &$18.987^{+.003}_{-.008}$ &\nodata &$285^{+47}_{-126}$ &$6.7\pm 1.4$ & $15.80^{+0.11}_{-0.13}$ &4.8 &\\
\ion{Ne}{9} K$\alpha$&13.447 &$13.451^{+.010}_{-.003}$ &\nodata &$89^{+223}_{-67}$ &$4.5\pm 1.1$ & $15.83\pm 0.21$ &4.1 &\\ 
\ion{N}{6} K$\alpha$ &28.787 &28.787 &\nodata &\nodata &$<8.4$ & $<15.39$ &$<2$ &\\
\ion{N}{7} K$\alpha$ &24.781 &24.781 &\nodata &\nodata &$<5.0$ & $<15.39$ &$<2$ &\\
\ion{O}{6} K$\alpha$ &22.019 &22.019 &\nodata &\nodata &$<5.7$ & $<15.36$ &$<2$ &\\
\ion{O}{7} K$\gamma$ &17.768 &17.768 &\nodata &\nodata &$<5.3$ & $<16.71$ &$<2$ &\\
\hline
UV (\fuse): \\
\hline\hline
\ovia &1031.926 &$1032.11\pm 0.01$ &$76.4\pm 4.1$ &$53.5\pm 2.9$ &$120.5\pm 3.8$ & $14.06\pm 0.02$ &31.7 &\\
\ovib &1031.926 &$1031.85\pm 0.01$ &$90.1\pm 8.1$  &$-22.1\pm 2.9$ &$98.2\pm 4.4$ & $13.94\pm 0.02$ &22.3 &\\
\hvca &1031.926 &$1031.48\pm 0.01$ &$74.2\pm 5.8$ &$-129.7\pm 2.9$ &$73.1\pm 4.0$ & $13.81\pm 0.03$ &18.3 &\\
\hvcb &1031.926 &$1031.12\pm 0.02$ &$80.3\pm 14.5$&$-234.3\pm 5.8$ &$46.6\pm 4.5$ & $13.59\pm 0.05$ &10.4 &\\
\enddata
\tablenotetext{a}{Rest wavelengths taken from \citet{verner96}, except
\ovi\ K$\alpha$ which is from the laboratory measurements of 
\citet{schmidt04}.}
\tablenotetext{b}{In the cases where upper limits were found, the line 
positions were allowed to vary within 20\,m\AA\ of the rest wavelengths,
approximately the RMS dispersion of the measured line wavelengths.
Measured wavelengths are taken from ACIS--S since its dispersion relation
is thought to have the fewest nonlinearities; however, only the statistical
fit error is given in this column (i.e., not taking into account 
systematic wavelength scale uncertainties of up to $\sim 20$\,m\AA).}
\tablenotetext{c}{Error bars are $1\sigma$; upper limits are $2\sigma$.}
\tablenotetext{d}{Column densities for X-ray lines are calculated assuming
the $1\sigma$ Doppler parameter region found in Figure~\ref{ch4_fig_nbovii} 
(35-94\kms\ for \ovii); for UV lines the measured $b$ values are used.}
\tablecomments{
(1) An upper error bar could not be formally determined for the wavelength
of this line.
}
\end{deluxetable*}

It should be emphasized that the observed wavelength errors reported in 
Table~\ref{ch4_tab_lines} are purely based on the fit to the data and do
not include systematic uncertainties in the wavelength scale.  At a minimum,
the nominal LETG wavelength error of $\sigma_\lambda \sim 10$\,m\AA\
should be assumed\footnote{See \url{http://cxc.harvard.edu/cal/}}.  
A more conservative estimate can be derived directly from the best--fit
positions of the six measured lines.  Assuming all of these ions are
at roughly the same velocity, the measured wavelengths exhibit an
RMS dispersion of $\sim 20$\,m\AA, similar to that seen in the Mrk 421
HRC spectrum \citep{nicastro07}.  Thus, the true
systematic uncertainty in the measured line positions (on top of the
fitted statistical errors) is on the order of $10-20$\,m\AA, consistent
with that found by the LETG calibration team$^2$.

\subsection{\fuse \label{ch4_sec_meas_fuse}}
The reduction and analysis of the \pks\ \fuse\ data were performed in a 
manner nearly identical to that described for the 
Mrk 421 \citep{williams05} and Mrk 279 \citep{williams06a} sightlines;
a brief summary follows.  Three observations of \pks\ totaling 120\,ks 
were available in the \fuse\ section of the Multimission Archive at
STScI website.\footnote{\url{http://archive.stsci.edu/}}  The calibrated
data were downloaded and individual exposures cross-correlated over the
1030--1040\,\AA\ range, where many strong absorption lines are present,
 to compensate for small (typically $<10$\,m\AA) 
variations in the wavelength scale.  The cross--correlated spectra for
each observation were then coadded, and the resulting spectra in turn
cross--correlated and coadded to produce a final coadded \fuse\ spectrum.
The absolute wavelength scale was checked by comparing the positions of the
strong, narrow
\ion{Si}{2} $\lambda 1020.699$ and \ion{Ar}{1} $\lambda 1048.220$ absorption 
lines to the \ion{H}{1} 21\,cm emission.  \citet{wakker03} find
that essentially all of the \ion{H}{1} is concentrated in a single Gaussian
component at $-4$\kms, which matches quite well the measured \ion{Ar}{1}
and \ion{Si}{2} velocities ($-5.4$ and $-5.9$\kms\ respectively, with
about 0.6\kms statistical error).  Since there may be small ($\sim$ few \kms)
systematic errors arising in the \ion{H}{1} measurement and/or the physical
relation between \ion{H}{1} and the two \fuse--measured species, we will
thus assume the \fuse\ wavelength calibration is correct for the purposes 
of this work.

\begin{figure}
\plotone{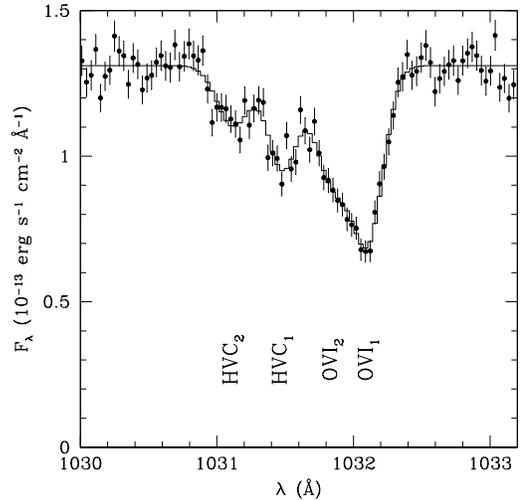}
\caption{\ovi\ $\lambda 1032$ region of the \pks\ \fuse\ spectrum.  The
positions of four Gaussian components used to model the $z\sim 0$ \ovi\ complex
are labeled.
\label{ch4_fig_fuse}}
\end{figure}

The \fuse\ spectrum shows strong \ovi\ $\lambda 1032$ absorption at
$v\sim 0$ as well as two distinct high--negative velocity \ovi\ 
components (hereafter referred to as \hvca\ and \hvcb\ in order of increasing
absolute velocity).  The 1029--1034\,\AA\ region of the spectrum was fit
with a constant continuum plus a single Gaussian for each of the three
\ovi\ components.  However, this provided a poor fit for the strong
low--velocity \ovi\ component so another Gaussian was added at
$v\sim 0$ to improve the fit.  Figure~\ref{ch4_fig_fuse} shows the resulting
data and best--fit model, and the parameters of the four Gaussian components
are listed in Table~\ref{ch4_tab_lines}.  Although \citet{wakker03} used a
different method to measure the equivalent widths of each \ovi\ component, 
our measured equivalent widths agree with theirs for both the
low-- and high--velocity \ovi\ components.

The velocity of the 
\hvcb\ component is inconsistent with the $z=0$ \ovii\ K$\alpha$
velocity at the $\sim 3\sigma$ level assuming the statistical error
on the line measurement, or $2.6\sigma$ if the nominal systematic
wavelength uncertainty of 10\,m\AA\ is adopted, indicating that the \ovii\
and \hvcb\ components may be kinematically distinct.  However, since 
wavelength scale errors in \chandra\ HRC--S/LETG are still not 
well--determined, this should not be considered a firm result.

The other \ovi\ doublet line at 1037.6\,\AA\ is also visible in the spectrum,
and in principle can be useful for curve--of--growth diagnostics when the
\ovi\ $\lambda 1032$ line is saturated.  With the high resolution of \fuse\
($\lambda/\Delta\lambda\sim 15000$), however, the 1032\,\AA\ line's shape
and strength has in the past been sufficient for these measurements.
Furthermore, the 1037\,\AA\ line components (particularly the HVCs) are 
heavily blended with nearby Galactic interstellar medium lines such
as \ion{C}{2}$^\ast$.  Since this blending can introduce additional
systematic error and even slight \ovi\ saturation appears to be rare
\citep[e.g.][]{wakker03}, we will disregard the 1037\,\AA\ \ovi\ line
in this analysis.

\section{Analysis}
\subsection{Doppler Parameters and Column Densities \label{ch4_sec_cog}}
The low resolution of the \chandra\ gratings compared to UV and optical
spectrographs presents unique challenges for column density measurements,
since essentially all non--quasar absorption lines are far narrower than the
50\,m\AA\ ($\sim 750$\kms\ at 20\,\AA) LETG line--spread function.  
The lack of line width information prevents direct measurement of the
profile shape, and hence the degree of saturation for any given line
cannot be directly determined.  If multiple absorption lines from the
same ionic species are detected, however, the relative equivalent
widths of these lines can instead be used to place limits on the
column density ($\novii$) and velocity dispersion (or Doppler
parameter, $b$) of the medium.

In the case of \pks, the \ovii\ K$\alpha$ and K$\beta$ lines are 
strongly detected, and an upper limit is measured for the K$\gamma$
line.  If all these lines were unsaturated, the equivalent widths would
scale as $W_\lambda\sim f_{ij}\lambda^2$ where $f_{ij}$ is the
absorption oscillator strength.  Saturation effectively decreases the
equivalent widths of strong (high--$f_{ij}$) lines while leaving 
weaker lines in the series more or less unaffected; thus, the ratio
$W_\lambda$(K$\beta$)/$W_\lambda$(K$\alpha$) increases with respect to
the simple (unsaturated) proportionality above.
For \ovii, the expected K$\beta$/K$\alpha$ equivalent width ratio
is $0.156$, while we measure a ratio of $0.36\pm 0.12$, indicating
that the \ovii\ K$\alpha$ may be slightly saturated (though it is also
consistent with no saturation at the $2\sigma$ level).

\begin{figure}
\plotone{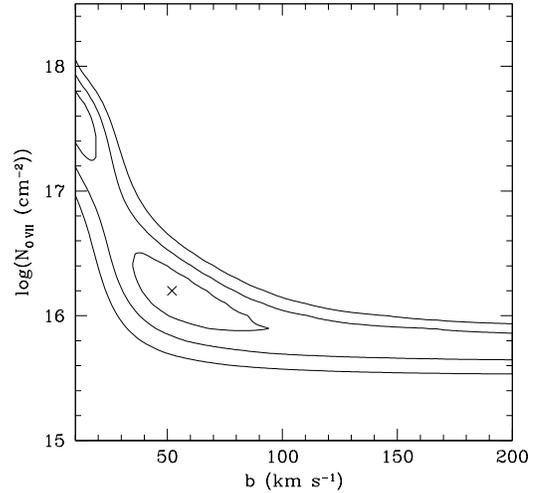}
\caption{\ovii\ Doppler parameter and velocity dispersion constraints
(at the $1\sigma$, $2\sigma$, and $3\sigma$ confidence levels) determined
by simultaneously fitting \ovii\ K$\alpha$, K$\beta$, and K$\gamma$ lines in the
joint LETG/ACIS+HRC \chandra\ spectrum.
\label{ch4_fig_nbovii}}
\end{figure}

To place more quantitative constraints on $\novii$ and $b$, we employ
the technique used in \citet{williams06a} for Mrk 279.  For a grid of
points in the $\novii-b$ plane, equivalent widths and apparent line
FWHM values were calculated numerically (using Voigt absorption line
profiles) for the \ovii\ K$\alpha$, K$\beta$, and K$\gamma$ transitions.
These lines were then added to the continuum model in \emph{Sherpa}, and 
$\chi^2$ calculated with the ``goodness'' command, for every value of
$\novii$ and $b$.  Figure~\ref{ch4_fig_nbovii} shows the contours of $1\sigma$,
$2\sigma$, and $3\sigma$ confidence calculated in this manner.

\begin{figure}
\plotone{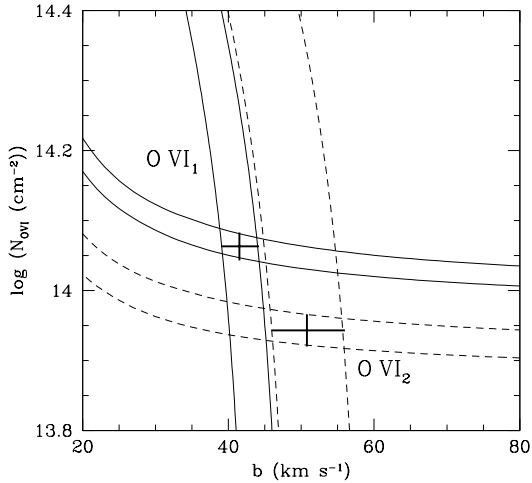}
\caption{Column density and velocity dispersion constraints derived from
\ovi\ $\lambda 1032$ equivalent widths (roughly horizontal lines) and
FWHMs (vertical lines), for the \ovia\ (solid) and \ovib\ (dashed) 
components.  Regions of overlap, marked with crosses, denote the approximate
$1\sigma$ confidence intervals on $\novi$ and $b$ for each component.
\label{ch4_fig_nbovi}}
\end{figure}

As this figure shows, the minimum $\chi^2$ is found at $b=52$\kms\ and
$\log(\novii)=16.2$, with the $1\sigma$ confidence region stretching between 
$b=35-94$\kms\ and $\log(\novii)=15.9-16.5$.  Additionally, another 
$1\sigma$ region can be found at $b<19$\kms\ with a higher column density
($\log(\novii)\sim 17.5$) required to produce consistency with the spectrum.
Such high \ovii\ column densities are unlikely to be produced in
a cold ($T_{\rm max}\la 3\times 10^5$\,K), weakly photoionized medium without
producing large amounts of narrow \ovi\ absorption, so a low--$b$ solution 
appears unlikely. 
However, it is important to note that no value of $b$ can be ruled out
at the $2\sigma$ confidence level from this curve--of--growth analysis
alone; as mentioned above, the absorption is
consistent with a completely unsaturated medium at this level, and lower--$b$,
higher--$\novii$ solutions are also possible in the regions demarcated by 
the $2\sigma$ and $3\sigma$ contours in Figure~\ref{ch4_fig_nbovii}.
Column densities for all X-ray lines (listed in Table~\ref{ch4_tab_lines}) 
are calculated by assuming the $\chi^2$--minimizing $b$ value above.  While
most of the lines are too weak for this choice to make a significant
difference in the $N_{\rm ion}$ determination, it should be kept in mind
that the systematic uncertainty in $\novii$ may be larger than the
statistical errorbars.

\begin{figure*}
\plotone{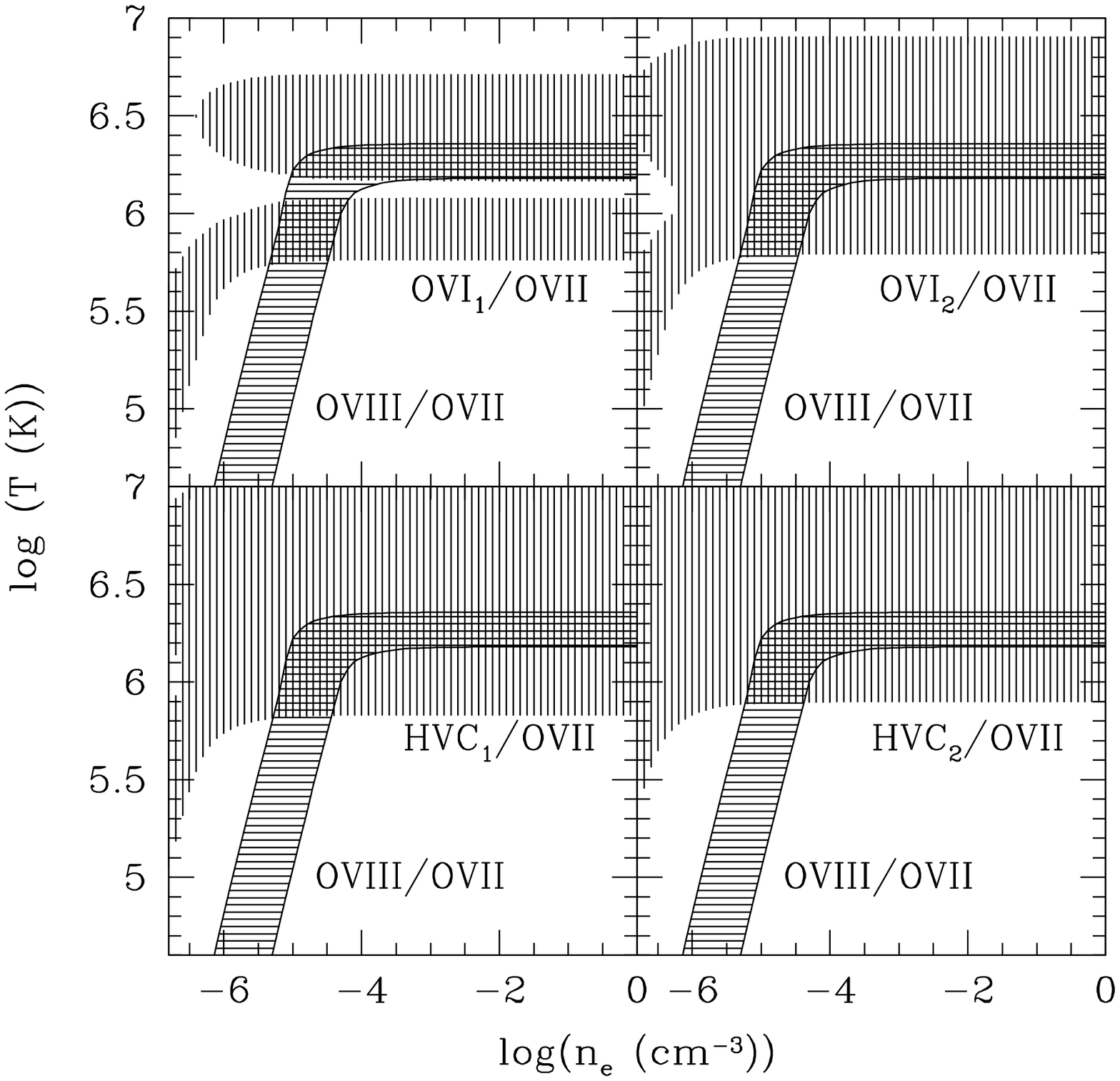}
\caption{Oxygen ion temperature and density constraints from $\noviii/\novii$
and $\novi/\novii$ for each of the four measured \ovi\ components.  
\label{ch4_fig_oxplot}}
\end{figure*}

Determination of these parameters for the UV \ovi\ $\lambda 1032$ absorption 
is decidedly more straightforward since the lines are fully resolved by
\fuse.  Since saturation can make absorption lines broader than would
be expected just from the Doppler parameter of the gas, the measured
line width cannot be used directly as a surrogate for $b$.  Instead,
we calculate \emph{apparent} \ovi\ FWHM values and equivalent widths
over a grid of $\novi$ and $b$, and find the regions within this grid
that are consistent with the measured $\Delta v_{\rm FWHM}$ and 
$W_\lambda$ values. 

Figure~\ref{ch4_fig_nbovi} shows these tracks for the low--velocity \ovi\ 
components.  Contours of constant $\Delta v_{\rm FWHM}$ are roughly
vertical while constant $W_\lambda$ are horizontal in the unsaturated
regime.  In this case both \ovia\ and \ovib\ appear to be at most
weakly saturated, so the $\Delta v_{\rm FWHM}$ and $W_\lambda$ contours
overlap nearly orthogonally, producing tight constraints on both parameters
for both components.  We find that $b=41.5\pm 2.5$\kms\ and 
$b=51\pm 5$\kms\ for \ovia\ and \ovib\ respectively, with column densities
of $\log(\novi)=14.06\pm 0.02$ and $13.94\pm 0.02$.  Since these low--velocity
lines are essentially unsaturated, and the HVCs are weaker still (but with
comparable apparent line widths), we can safely assume they fall well 
within the linear part of the curve--of--growth.  Values of $b=44.6\pm 3.5$
and $\novi=13.81\pm 0.03$ (for \hvca) and $b=48.2\pm 8.7$ and 
$\novi=13.59\pm 0.05$ (for \hvcb) are thus inferred directly from the line
measurements.  All four \ovi\ components have Doppler parameters
that are fully consistent with the \ovii\ $1\sigma$ limits.

\subsection{Temperature and Density Diagnostics \label{ch4_sec_tdiag}}
With estimates for ionic column densities, constraints on the temperature
and density of the absorbing medium can be derived.  
Although collisional ionization is expected to be the dominant physical
process in either the extended local WHIM or a hot Galactic corona, 
photoionization from the extragalactic UV/X-ray background is expected
to significantly alter the ionization balance of the low--density
WHIM \citep[cf.][]{nicastro02,mathur03}.  To find the most general set of 
conditions
which can produce the observed highly--ionized ion ratios, both collisional
and photo--ionization must be considered.

For this sightline we follow the same analysis we employed for Mrk 279
in \citet{williams06a}.
Assuming a fixed $z=0$ metagalactic ionizing background model from 
\citet{sternberg02}, the ionization parameter 
$U=n_\gamma(E>13.6{\rm eV})/n_e$ simply depends on the inverse of the electron
density.  The ionization balance code Cloudy 
\citep[version 05.04;][]{ferland98} was employed to calculate relative
abundances of all measured ions over a range of $T=10^{4.5}-10^{7.4}$\,K
and $n_e=10^{-7}-1$\,cm$^{-3}$ (or $U=10^{0.7}-10^{-6.3}$) with grid
spacings of 0.1\,dex in each quantity, encompassing the range
of temperatures and densities expected in WHIM and Galactic corona models.

With a grid of $N_i$ computed as a function of temperature and density, the 
problem can be inverted to determine which sets of $T$ and $n_e$ are consistent
with the measured ionic column densities.  However, since the local X-ray
absorption is produced in gas too hot to be detectable in neutral hydrogen
emission (and Ly$\alpha$ absorption at $v\sim 0$ is invariably obliterated 
by the local interstellar medium damping wing), no information on the
overall metallicity can be derived from the data.  Thus, it is more useful
to find the $\log T-\log n_e$ regions defined by column density 
\emph{ratios}.  Since \ovii\ is by far the best--measured ion that 
unambiguously arises in local warm--hot gas, we calculate all other ion
column density ratios relative to $\novii$.

Column density ratios of different ions of the same element are independent
of metallicity, and so depend only on the physical state of the medium.
Thus, if the \oviii\ and \ovii\ absorption arise in the same gas phase,
the $\noviii/\novii$ ratio provides the most rigorous constraints on the
temperature and density of the warm--hot gas.  Likewise, if any one of the
four measured \ovi\ components exists in this same phase, the $\novi/\novii$
ratio should be consistent with an overlapping set of temperatures and
densities.  Figure~\ref{ch4_fig_oxplot} shows the $2\sigma$ constraints derived
from $\noviii/\novii$ and $\novi/\novii$ for each of the \ovi\ components.
Note that at high densities ($n_e\ga 10^{-4}$\,cm$^{-3}$) the temperature
constraints are essentially constant, but at lower densities photoionization
becomes significant and a lower temperature is necessary to produce the
same column density ratios.  

The $\log T$ and $\log n_e$ values derived
from the measured $\noviii/\novii$ are consistent with the $\novi/\novii$
constraints for all four components, with a typical minimum density of
$n_e\ga 10^{-5}$\,cm$^{-3}$.  However, the consistency
for the strongest low--velocity \ovi\ component (\ovia) is only marginal
($\sim 2\sigma$ level).  Since \ovia\ and \ovib\ are almost certainly
associated with the Galactic thick disk, the density of these media must
be high enough that only collisional processes affect the ionization balance.
The collisional--equilibrium temperature 
inferred from the \oviii/\ovii\ ratio cannot be reconciled with large
amounts of \ovi\ in the same phase ($\log(\novi)\ga 14.1$), so the sum of 
all \ovi\ 
components associated with the \oviii/\ovii\ absorber must have a total
column density lower than this value.
Thus, association of the X-ray oxygen
absorption lines with any one of the UV \ovi\ components cannot be ruled
out for this sightline, but the X-ray absorption cannot be associated
with multiple strong \ovi\ components.  If the medium is collisionally 
ionized, the 
$\noviii/\novii$ ratio provides strong $2\sigma$ temperature constraints
(assuming a Doppler parameter $b\sim 52$\kms) of $6.18<\log T<6.36$.
If the density is low enough ($n_e\sim 10^{-5}$\,cm$^{-3}$) that 
photoionization is important, then the temperature can be as low as
$T\sim 10^{5.5}$\,K.

\begin{figure*}
\plottwo{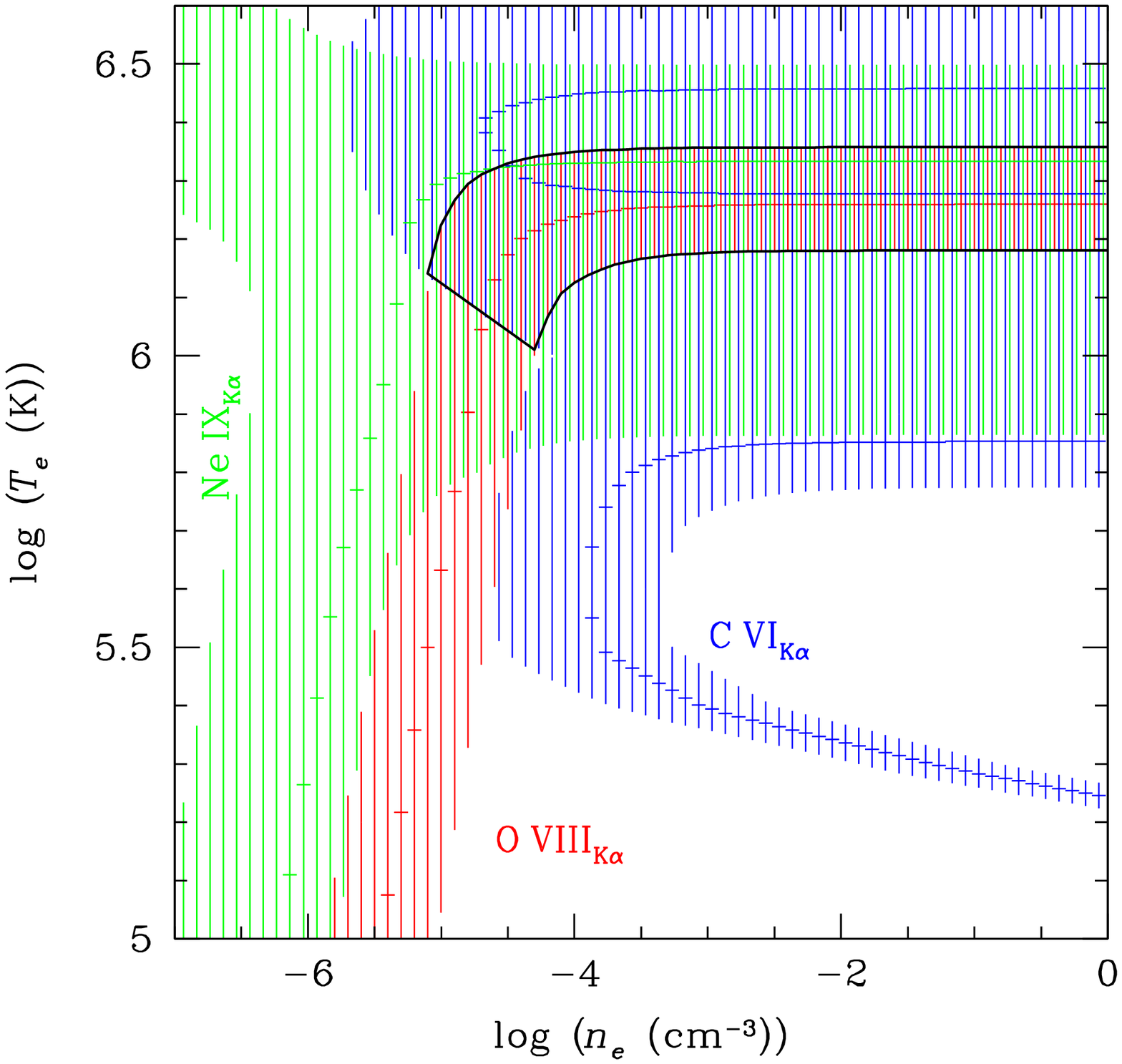}{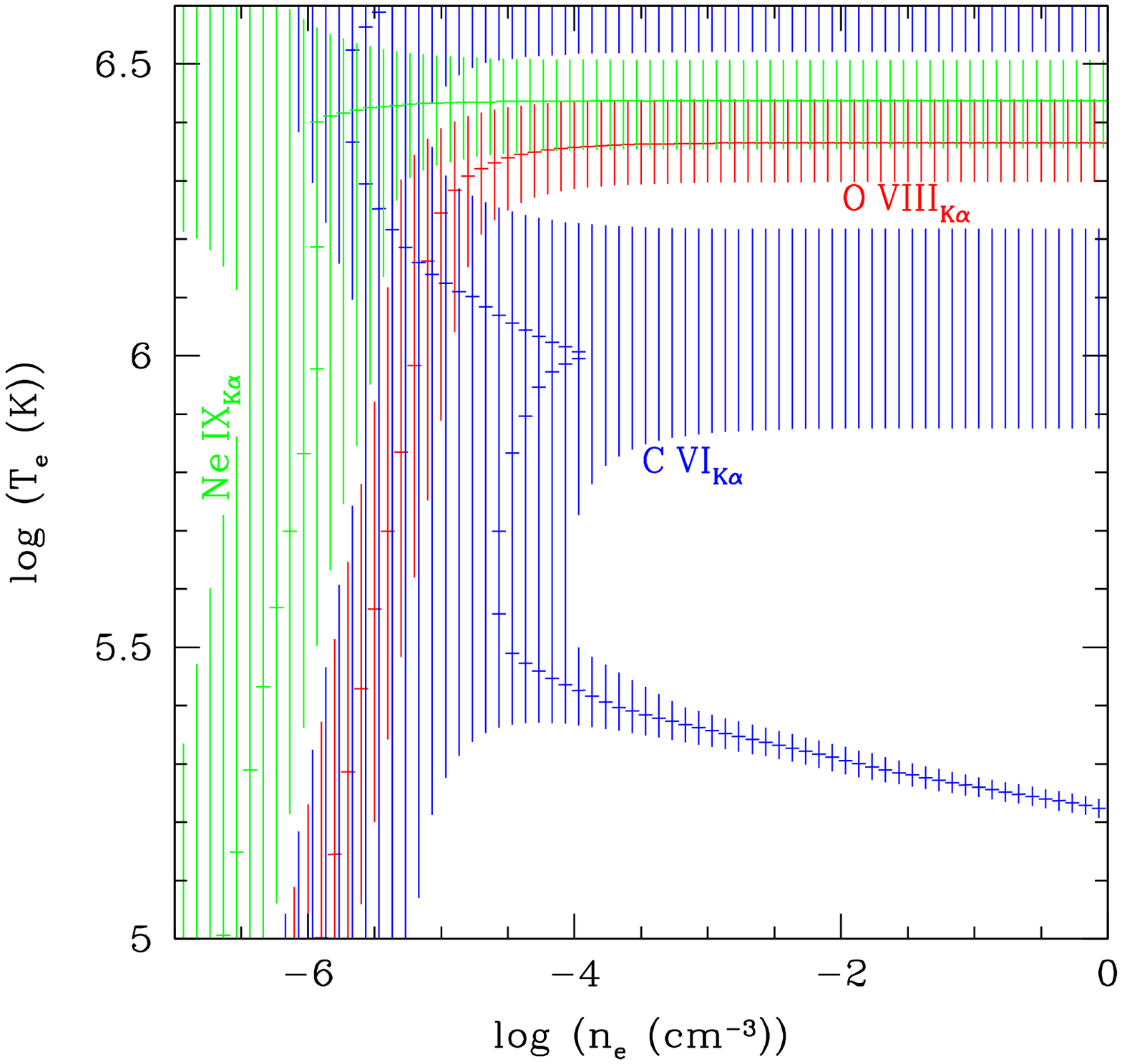}
\caption{\emph{Left panel:} Temperature and density constraints from ion 
column density ratios
relative to $\novii$, where column densities are derived assuming the 
best--fit region near $b=50$\kms\ shown in Figure~\ref{ch4_fig_nbovii}. The
dark black line shows the approximate region of overlap between the three 
contours (defined primarily by $\noviii/\novii$ and $N_{\rm CVI}/\novii$);
\emph{Right panel:} Same as left, but with column densities derived assuming
negligible saturation ($b=200$\kms).  There is essentially no combination
of $n$ and $T$ consistent with all three ion ratios in this case.
\label{ch4_fig_ntb52}}
\end{figure*}

Similar constraints can be derived from other measured X-ray lines, though
since (for example) the ratio $N_{\rm CVI}/\novii$ depends on the relative [C/O]
abundance, these constraints are more prone to systematics
than those solely employing one atomic species.  Moreover, since the 
\ovii\ Doppler parameter cannot be pinned down accurately,
$\novii$ can vary with different $b$ values thus significantly altering
the measured column density ratios.  For the low--$b$ solution ($b<19$\kms),
the measured $\noviii/\novii$ requires temperatures of $\log T\sim 6.2$.
On the other hand, such a low velocity dispersion implies a maximum temperature
of $\log T_{\rm max}= 5.5$ so the low--$b$ solution does not appear to be
physically possible.  For these X-ray line diagnostics, we thus consider only 
the best--fit $1\sigma$ region of $b=52^{+42}_{-35}$\kms\ and a large--velocity
dispersion, low saturation ($b\sim 200$\kms) solution.
 
Figure~\ref{ch4_fig_ntb52} shows the constraints
for $b=52$ and $b=200$ derived from the ratios of \neix, 
\cvi, and \oviii\ to \ovii, assuming a solar abundance pattern for these
elements.  The column density limits measured for
\nvi\ and \nvii\ did not provide any useful constraints (i.e., they were
consistent with nearly the entire range of temperatures and densities) 
and were excluded from these figures for clarity.  Similarly, the
\cv\ ion is expected to form in cooler gas than that producing the \ovii\ 
and \oviii\ absorption and may be contaminated by foreground ISM or thick-disk
gas, so the $N_{\rm CV}/\novii$ contour (which provides no additional
constraints) is also not shown.

In the best--fit $b$ plot (Figure~\ref{ch4_fig_ntb52}, left panel), the constraints 
derived from all three column density ratios overlap quite well in the
collisionally--ionized density regime ($n_e\ga 10^{-5}$\,cm$^{-3}$).  
On the other hand, at a large velocity dispersion ($b=200$; right panel of Figure~\ref{ch4_fig_ntb52})
there is essentially no set of $\log T$ and $\log n_e$ for which the
three constraints overlap.  The \cvi\ and \neix\ lines are relatively
weak and not as affected by saturation as the \ovii, so this change is
driven primarily by the decrease in $\novii$ at higher velocity dispersion.
Though the contours derived from $N_{\rm CVI}$ 
and $N_{\rm NeIX}$ depend on [C/O] and [Ne/O], if the abundance 
mixture of this absorber is roughly Solar, then the measured column
densities indicate that the previously derived Doppler parameter range
($b=35-94$\kms) fit the data better than a high--$b$, unsaturated medium.

\citet{collins04} perform a similar analysis on the UV/X-ray absorption
toward this sightline, employing the \citet{nicastro02} X-ray measurements
and assuming (as a maximal case) that both \ovi\ HVCs are associated with
the X-ray absorption.  They find that the 
\ovi\ column density seen in the HVCs is far larger than that expected
from the \ovii\ column density and high temperature 
($\log T=6.30\pm 0.15$) inferred from 
$\noviii/\novii$.  However, as discussed below in \S\ref{sec_nicastro}, 
the improved constraints on $b$ imply larger \ovii\ column densities
(hence lower \ovi/\ovii) than found in \citet{nicastro02}, so the large 
observed \ovi\ HVC column densities now are consistent with the \ovii\ 
and \oviii.  The scenario put forth by \citet{collins04} (\ovi\ arising
in a warm--hot collisionally--ionized medium, with lower ions coexisting
in a cooler, photoionized phase) thus remains consistent with the data, but 
the \ovii\ and \oviii\ can now be included in the warm--hot phase as well.

\subsection{Absorption at $z=0.055$}
In a previous study incorporating a subset of the data analyzed herein
(ACIS--S/LETG observation IDs 1703, 2335, and 3168), 
\citet[][hereafter F02]{fang02}
report the detection of an absorption line at $20.02\pm 0.015$\kms\ with
equivalent width $14.0^{+7.3}_{-5.6}$\,m\AA,
possibly corresponding to \oviii\ at a velocity of $16634\pm 237$\kms\
($z=0.055\pm 0.001$).  Several previously discovered Ly$\alpha$ absorption
lines and a small cluster of \ion{H}{1} galaxies appear at a similar
velocity in this direction \citep{shull98}, so such an absorber may be 
indicative of intragroup medium or an associated large--scale WHIM filament.

\begin{figure}
\plotone{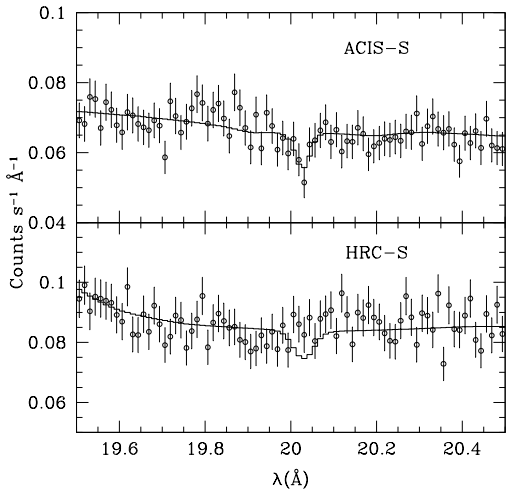}
\caption{ACIS--S and HRC--S/LETG (upper and lower panels, respectively) 
spectra of \pks\ near the wavelength of the $z=0.055$ \oviii\ intervening
feature reported by \citet{fang02}.  The lack of a detection in the HRC
spectrum does not rule out the ACIS detection.
\label{ch4_fig_z055spec}}
\end{figure}

This absorption line is clearly visible in our coadded ACIS spectrum, 
but is not visible in the HRC spectrum (Figure~\ref{ch4_fig_z055spec}).  
A fit to the line in ACIS
yields a wavelength of $20.03\pm 0.01$\,\AA\ and an equivalent width
$W_\lambda=7.5\pm 2.1$\,m\AA\ ($\log\noviii =15.76^{+0.12}_{-0.16}$ assuming
$b\sim 100$\kms), both consistent with the F02
measurement (though the best--fit equivalent width from F02 is roughly
twice that measured here).  Note that this line falls near an ACIS--S chip
node boundary, potentially introducing a systematic offset to the line 
strength, but such an offset is expected to be small compared to the 
measurement error.
An upper limit for a line at this wavelength ($\pm 0.02$\,\AA,
to account for possible HRC/LETG wavelength scale discrepancies)
in the HRC spectrum
was calculated and found to be $W_\lambda<12.5$\,m\AA\ ($2\sigma$ confidence),
so the ACIS measurement is not ruled out by the HRC non--detection.  
A simultaneous fit to both the ACIS and HRC spectra produces
almost exactly the same equivalent width and errors as the ACIS data alone,
further demonstrating that the HRC spectrum is insensitive to this line.
If real, this detection
would still be fully consistent (albeit with large errors) with the 
predicted number of \oviii\ absorbers per unit redshift (F02, Figure 2).

Even with the increased signal--to--noise provided by the
additional observations, no other absorption lines are seen at this
redshift, despite \ovii\ absorption often being present at WHIM temperatures.
From the $2\sigma$ upper limit
of $\log\novii<15.3$ at this redshift and the measured \oviii\ column
density of $\log\noviii =15.76^{+0.12}_{-0.16}$ (assuming $b\sim 100$\kms),
the inferred $2\sigma$ \emph{lower} limit on the absorber temperature
is $\log T>6.4$ in collisional ionization equilibrium.  This is larger
than the maximum temperatures inferred for local warm--hot gas
\citep[e.g., the Galactic corona
or Local Group; see][and references therein]{williams06a}, but
it falls within the predicted temperature range for the WHIM.
Furthermore, at densities
below $n_e\la 10^{-5}$\,cm$^{-3}$ photoionization enhances the \oviii\
column density in the medium, and
the corresponding temperature limit relaxes somewhat (e.g.~
$\log T>5$ for $n_e=10^{-6}$\,cm$^{-3}$).

\section{Discussion}
\subsection{Comparison to Other Lines of Sight}
\pks\ represents the third \chandra\ data set we have analyzed for 
which relatively weak ($\sim 10$\,m\AA) local absorption, uncontaminated by
known hot foreground gas (as in 3C 273, situated along the line of sight
to a likely Galactic supernova remnant; Savage et al.~1993) 
can be detected at high confidence.  Although 
the number of such sightlines is small, similarities and differences in
the detected absorption are already beginning to emerge.

\subsubsection{Mrk 421}
The \chandra\ LETG spectrum of Mrk 421 contains is the highest--quality
grating spectrum of an extragalactic source to date, with roughly three
times the counts per resolution element at 21\,\AA\ as the \pks\ data
analyzed here \citep{nicastro05a,williams05}.  Many of the ionic species
seen toward Mrk 421 (particularly \ovii, \oviii, \neix, and \cvi) were
also detected in the \pks\ spectrum, allowing direct comparisons to be
made.  In particular, the \ovii\ absorption (the strongest--detected
ion in both cases) shows strikingly similar properties between the two
objects--$\novii=16.23\pm 0.21$ in Mrk 421 versus $16.09\pm 0.19$ 
in \pks.  Both \ovii\ absorbers exhibit consistent velocity
dispersions as well, with $b_{\rm OVII}=24-55$\kms\ ($2\sigma$ limits)
in Mrk 421 and $35-94$\kms\ in \pks\ (though this latter quantity is
the $1\sigma$ range, and $2\sigma$ limits could not be determined).

One of the most surprising features of the Mrk 421 \chandra\ spectrum
is the presence of a weak (2.0\,m\AA) absorption line at 22.02\,\AA, 
the expected wavelength of the \ovi\ K$\alpha$ inner--shell transition.
Even though both this transition and the 1032\,\AA\ lines should both
trace the \ovi\ ground state, the column density derived from the
observed K$\alpha$ line is a factor of $\sim 4$ higher than that derived
from the UV transition.  If this K$\alpha$ transition is a better tracer
of the true \ovi\ column density than the UV line, then the \ovii\ 
toward Mrk 421 almost certainly arises in a low--density, photoionized
medium.

Although the \ovi\ $\lambda 1032$ absorption strength toward \pks\ is
comparable to the Mrk 421 sightline, unfortunately this \chandra\ 
spectrum does not have sufficient signal--to--noise to detect the
K$\alpha$ line.  Thus, for \pks\ we cannot determine if there is the
same discrepancy between the UV and X-ray \ovi, or if the inferred
\ovi\ K$\alpha$ column density also implies a photoionized medium along
this sightline.  If the \ovi\ K$\alpha$ measurement is disregarded, then
the Mrk 421 spectrum implies slightly lower--but marginally
consistent--temperatures in the collisionally
ionized regime ($\log T=6.1-6.2$; $2\sigma$ limits from the \oviii/\ovii\
ratio) than \pks\ ($\log T=6.18-6.36$).  The lower density limit from Mrk
421 is somewhat more stringent than that derived from the \pks\ 
oxygen ion ratios ($\log n_e> -4.7$ versus $>-5.5$), though this may
again be primarily an issue of spectral quality.

\subsubsection{Mrk 279}
The $z=0.03$ Seyfert galaxy Mrk 279 is significantly less luminous than
either \pks\ or Mrk 421, but it was observed long enough with
\chandra\ HRC--S/LETG to produce a reasonably high--quality spectrum
in which strong $z=0$ \ovii\ K$\alpha$ absorption was detected
\citep{williams06a}.  Two features of this absorption were particularly
interesting: (1) the unusually strong \ovii\ K$\alpha$ absorption 
($W_\lambda=26.6\pm 6.2$\,m\AA), coupled with a tight upper limit on 
the K$\beta$ line, indicated that the absorption was best described
as an unsaturated medium (with a $2\sigma$ lower limit of $b>77$\kms\
on the Doppler parameter); and (2) the \ovii\ absorption appears slightly
redshifted, making its velocity inconsistent with the \ovi\ HVC's negative
velocity at the $2.5\sigma$ level.  

As previously mentioned, the velocity of the \ovii\ toward \pks\ appears
to be inconsistent with the \hvcb\ velocity at the $2.6\sigma$ level, 
assuming the nominal HRC--S/LETG wavelength scale uncertainty of 10\,m\AA.
Until systematic errors in the wavelength scale can be better understood,
however, this should not be taken as a definitive result.  Moreover,
while the \ovii\ toward Mrk 279 could not be directly associated with
either low--velocity \ovi\ component since their velocity dispersions
were significantly different, such an association cannot be ruled out
in the \pks\ absorption: all of the \ovi\ Doppler parameters fall within the
$1\sigma$ $b$ confidence interval found for the \ovii\ absorption.

Temperature and density constraints on the $z=0$ absorption 
toward Mrk 279 and \pks\ are consistent with each other, though this
is not surprising--since the quality of the Mrk 279 spectrum is lower, 
only an upper limit on temperature and a lower limit on density could
be derived.

\subsection{Where is the Absorption?}
In principle, the degree of photoionization in an absorbing medium
(and hence an estimate of the gas density) can be derived from ionic
column density ratios (\S\ref{ch4_sec_tdiag}).  However, in the case
of \pks, the errors are large enough that no upper limit on the gas
density can be found, i.e.~it is fully consistent with collisional 
ionization or a combination of collisional-- and photo--ionization.  Depending 
on the assumptions made (in particular, which
if any of the \ovi\ components are associated with the \ovii\ 
and \oviii), the minimum density of this medium appears to be
$\log n_e \ga -5.5$.  The best--fit \ovii\ column density
is $\log \novii=16.09$ and \ovii\ is by far the dominant ionization
state in this medium.  If the gas has a metallicity of
$0.3\times$ solar (comparable to that observed in the diffuse intracluster
medium), then the total hydrogen density is roughly $\log N_{\rm H}\sim 19.9$
and the thickness of the absorber $d\la 10^{25.4}\;{\rm cm}=8.4$\,Mpc.

Thus, under a set of reasonable assumptions, the observed absorption is
consistent with an extended extragalactic medium, but within the large
uncertainties it is just as reasonable
to associate it with a local hot Galactic corona.  Although an association
between the X-ray absorber and low--velocity \ovi\ cannot be ruled out 
from these data alone, the properties of the X-ray absorber are quite
similar (again within the errors) to those studied along the Mrk 421
and Mrk 279 sightlines.  These latter two X-ray absorbers are
definitely not associated with the low--velocity \ovi\ absorption arising
in the Galactic thick disk, so if they indicate the presence of an
additional hot Galactic component, then the derived properties of the
\pks\ absorber are not in conflict with the other measurements of this 
component.

\subsection{Comparison to \citet{nicastro02} \label{sec_nicastro}}
In their study of three HRC--S/LETG observations of \pks\ (observation
IDs 331, 1013, and 1704), \citet[][hereafter N02]{nicastro02} also 
detected \ovii\ K$\alpha$
and K$\beta$, \oviii, and \neix, albeit at lower confidence.  They also
analyze \fuse\ data of the same sightline, but at that time only 39\,ks
were available, or about one--third of the exposure time analyzed here.
As it turns out, the addition of new \chandra\ and \fuse\ data
brings about significant changes in the interpretation of the 
local absorption, in two important ways.  

First, while N02 were able to fit the observed
\ovi\ $\lambda 1032$ line with two Gaussian components (one low--velocity
narrow line and a broader, blueshifted HVC), the new higher--quality spectrum
reveals that the ``broad'' component is actually two distinct HVCs, and
the ``narrow'' low--velocity \ovi\ is best fit with two components.  Second,
while the \ovii\ K$\beta$ line was previously not detected strongly enough
to place constraints on the Doppler parameter of the absorption, here the
K$\alpha$ and K$\beta$ lines hint at some degree of saturation and so
the \ovii\ column density we employ in our analysis is about 0.5\,dex 
higher than that reported by N02.  Both of these effects cause a sharp
decrease in the $\novi/\novii$ ratio, which in turn removes the need
for a photoionization contribution.  Indeed, when we calculate
temperature and density constraints assuming $b=200$ (as N02 had done;
see Figure~\ref{ch4_fig_ntb52}, right panel) we also find that a high--density,
collisionally--ionized solution cannot be found without modifications
to the relative abundances.  This highlights the major improvements in
diagnostic power that can be made by accumulating large numbers of counts
per resolution element, either through very long exposures or observations
of especially bright background sources.

\section{Conclusions}
Using all available \chandra\ LETG data on \pks, we have analyzed in
detail the ionization and kinematic state of the warm--hot $z=0$
absorbing medium.  We find a Doppler parameter range of $b=35-94$\kms\
($1\sigma$ limits; $2\sigma$ limits could not be found), which is
consistent with the absorption seen toward both Mrk 421 and Mrk 279
(though the best--fit value best matches the former sightline).  
Assuming that the Doppler parameter lies in this range, ionic column
densities of \ovii, \oviii, \neix, and \cvi\ are consistent with 
collisional ionization at $\log T({\rm K})=6.18-6.36$, though a
low--density WHIM with a significant photoionization contribution
cannot be ruled out.  Unlike the other
two previously analyzed sightlines, the \ovii\ absorption toward
\pks\ may be associated with either one of the low--velocity \ovi\ 
$\lambda 1032$ components seen in \fuse\ or a high--velocity \ovi\ 
cloud at $v=-130$\kms\ (though its velocity may be inconsistent at the
$\sim 2.5\sigma$ level with another \ovi\ HVC at $v=-234$\kms, and it is
only marginally consistent with the column density of the strongest
low--velocity \ovi\ component).  The similarities and differences between
different sightlines suggest that perhaps there is no single solution
to the origin of the $z=0$ \ovii\ absorbers.  The relation between \ovi\
HVCs and \ovii\ also appears to be diverse, so the location of these systems
remains an open question.

The intervening \oviii\ absorber at $z=0.055$ reported by \citet{fang02}
is detected in ACIS and is consistent with their measurement; although
no other lines are detected at the same redshift, this \oviii\ absorption is
consistent with expectations for WHIM gas.

It is notable that the \chandra\ data for this line of sight hint at
a number of interesting results (especially the low best--fit Doppler
parameter in the $z=0$ absorption) but the data are not quite of sufficient 
quality to confidently
confirm them.  \pks\ is quite possibly the only other extragalactic source 
bright enough
to obtain a \chandra\ LETG spectrum with $\sim 6000$ counts per resolution
element, comparable to the Mrk 421 spectrum analyzed by \citet{williams05},
in a reasonable amount of time.  Such a spectrum would not only allow
a direct comparison of the $z\sim 0$  absorption along two lines of
sight, but would also provide a path length four times
larger than Mrk 421 to search for ``missing baryons'' in intervening WHIM 
filaments.  With two such systems detected in Mrk 421, a correspondingly
larger number could be detected in a \pks\ spectrum at comparable
signal--to--noise.  While longer observations of PKS~2155 would have the 
potential for great scientific results, the large soft X-ray variability 
of this source (over a factor of ten, as seen in Table~1) could hamper
the feasibility of such observations.

\acknowledgments

The authors thank the \chandra\ and \fuse\ teams for their efforts on these
superb missions.
Ionization balance calculations were performed with 
version 05.07 of Cloudy, last described by \citet{ferland98}.
This research has been supported by \chandra\ award AR5-6017X issued by the
\chandra\ X-ray Observatory Center, which is operated by the Smithsonian
Astrophysical Observatory for and on behalf of the NASA under contract
NAS8-39073.  F.~N.~acknowledges support from NASA LTSA grant
NNG04GD49G.

\end{document}